\documentclass[fleqn,12pt,twoside]{article}
\usepackage{espcrc1}
\input{epsf}

\usepackage{graphicx}

\newcommand{\Nmassmed}{m_{N}^*} 
\newcommand{\Nmassvac}{m_{N}} 
\newcommand{\Nstarmassmed}{m_{N^*}^*} 
\newcommand{\Nstarmassvac}{m_{N^*}} 
\newcommand{\Nstarwidmed}{\Gamma_{N^*}} 
\newcommand{\etamass}{m_\eta}  
\newcommand{\getaNR}{g_\eta} 
\newcommand{\mevsig}{\langle \sigma \rangle} 

\title{Medium effects to the $N(1535)$ resonance and $\eta$ mesic
nuclei}

\author{D.\ Jido\address[RCNP]{Research Center for Nuclear Physics, 
    Osaka University,
    Ibaraki, Osaka 567-0047, Japan},
        H.\ Nagahiro\address[NWU]{Department of Physics, 
     Nara Women's University, Nara 630-8506, Japan}
        and
        S. Hirenzaki\addressmark}

\begin{document}

\maketitle

\begin{abstract}
The structure of $\eta$-nucleus bound systems ($\eta$ mesic
nuclei) is investigated as one of the tools to study in-medium properties of the
$N(1535)$ ($N^{*}$) resonance.  We show that, as a general consequence, 
the $\eta$-nucleus potential has a repulsive core
at the nuclear center with an attractive part at the
nuclear surface, if sufficient reduction of the mass difference of $N$ and $N^*$
stems from the in-medium effects to $N^*$.  
The (d,$^3$He) spectra are evaluated for the formation 
of these bound states to investigate the experimental feasibility.
\end{abstract}

\section{Introduction}

The study of the in-medium properties of hadrons has attracted continuous
attention and is one of the most interesting topics of
nuclear physics.  
In the contemporary point of view, in-medium properties of hadron
are believed to be related to partial restoration of chiral symmetry,
in which a reduction of the order
parameter of the chiral phase transition in hot and/or dense matter takes 
place and causes modifications of the hadron properties.

In this paper we consider the $\eta$ mesic nucleus as one of the doorways to 
investigate the in-medium properties of the $N(1535)$ ($N^*$).
The special features of the $\eta$ mesic nucleus are the following; 
(1) the $\eta$-$N$ system dominantly couples to the $N^*$ at the
threshold region.
(2) The isoscalar particle $\eta$ filters out contaminations of the
isospin 3/2 excitations in the nuclear medium.
(3) Due to the $s$-wave nature of the $\eta NN^*$ coupling there is no
threshold suppression like the $p$-wave coupling.  
The strong coupling of the $N^*$ to $\eta N$ makes the use of this channel 
particularly suited to investigate this resonance in a cleaner way than the use of 
$\pi N$ for the study of other resonances like the $N(1440)$ and $N(1520)$.

The  $N(1535)$, which is the
lowest lying parity partner of the nucleon, has been investigated from the
point of view of chiral symmetry \cite{DeTar}.
Considering the fact that the $N^*$ mass in free space lies only 50
MeV above the $ \eta N$ threshold, the medium modification of
the $N^*$ mass will strongly affect the in-medium potential of the
$\eta$ meson through the strong $\eta NN^{*}$ coupling mentioned  
above.

\section{Optical potential of $\eta$ with $N^*$ dominance}
The $\eta$-mesic nuclei were
studied by Haider and Liu \cite{haider86} and by Chiang, Oset and Liu
\cite{chiang91} systematically.  There, the $\eta$-nucleus optical
potential was expected to be attractive from the data of the
$\eta$-nucleon scattering length and the existence of the bound states
was predicted theoretically.

First of all, we show, as a general consequence, 
the possibility to have a repulsive $\eta$ optical potential in the nucleus due to 
a significant reduction of the mass difference of $N$ and $N^*$.  Considering the
self-energy of the $\eta$ meson at rest in nuclear matter in the
$N^*$ dominance model, in analogy with the $\Delta$-hole model for
the $\pi$-nucleus system, we obtain the $\eta$ optical potential  in the 
nuclear medium in the heavy baryon limit \cite{chiang91} as;
\begin{equation}
   V_\eta(\rho(r); \omega) = {\getaNR^2 \over 2 \mu} 
   {\rho(r) \over \omega + \Nmassmed(\rho)  - \Nstarmassmed(\rho) + i
   \Nstarwidmed(s;\rho)/2} \ , \label{poteta}
\end{equation}
where $\omega$ denotes the $\eta$ energy, and $\mu$ is the reduced mass
of the $\eta$ and  the nucleus. The nucleon density distribution $\rho(r)$ is 
assumed to be a Fermi distribution 
in the finite nucleus.
The ``effective mass'' of $N$ and $N^*$ in medium are
denoted as $\Nmassmed$ and $\Nstarmassmed$. 
The in-medium $N^*$ width $\Nstarwidmed$ includes the many-body
decay channels.  The $\eta NN^*$ vertex is taken as the 
isoscalar and scalar coupling with $g_\eta \simeq 2$.

Let us suppose no medium modifications for the masses  of  $N$ and $N^*$.
This is nothing but the $T\rho$ approximation.
In the case of 
small binding energy for the $\eta$, {\it i.e.} $\omega \simeq \etamass$,
we obtain an attractive potential independent of density because of
$\omega + \Nmassvac - \Nstarmassvac < 0$.  In this case, the shape
of this potential is essentially the same as the Woods-Saxon type potential for 
a finite nucleus.  On the
other hand, if a sufficient reduction of the mass difference of $N$
and $N^*$ stems from the medium effects, there exists a critical density
$\rho_c$ where $\omega + \Nmassmed - \Nstarmassmed = 0$, and then at
densities above $\rho_c$ the $\eta$ optical potential turns to be
repulsive.  If $\rho_c$ is lower than the nuclear saturation
density $\rho_0$, the optical potential for the $\eta$ is  attractive
around the surface of the nucleus and repulsive in the interior.


To make the argument more quantitative, we estimate the in-medium $N$
and $N^*$ masses and the $N^*$ width in the chiral doublet model
\cite{DeTar}, which is an extension of the
$SU(2)$ linear sigma model for the nucleon incorporating the $N^*$ 
in a chiral symmetric way. The chiral doublet model represents the 
mass difference of $N$ and $N^*$ as a linear function of the 
chiral condensate, and gives the density dependence of the mass difference 
of $N$ and $N^*$ in the mean-field approximation \cite{Kim} as
\begin{equation}
   \Nmassmed (\rho) -\Nstarmassmed (\rho)= (1-C {\rho/\rho_0})(\Nmassvac -
   \Nstarmassvac) \label{massdiff} \ ,
\end{equation}
where we take a linear 
parameterization of the in-medium chiral condensate, 
$  \mevsig = (1-C\rho/\rho_0) \langle \sigma \rangle_0 $, 
with $C$=$0.1$-$0.3$ \cite{HKS}. 
The $C$ parameter represents the strength of 
the chiral restoration at 
the nuclear saturation density $\rho_0$.

The medium effects on the decay width of $N^*$ are 
taken into account by considering the Pauli blocking effect on 
the decaying nucleon,  by changing the $N$ and $N^*$ 
masses and the $\pi NN^*$ coupling in medium
according to the chiral doublet model, and also by considering 
the many-body decays of $N^*$, 
such as $N^*N \rightarrow NN$ and $N^*N \rightarrow \pi NN$. 

In the present calculation, the chiral doublet model with the mirror 
assignment is used.
The detail of this work is discussed in ref.\cite{Jido:2002yb}.

\section{Numerical Results}

\begin{figure}[tb]
\epsfxsize=14.5cm
\begin{center}
\epsfbox{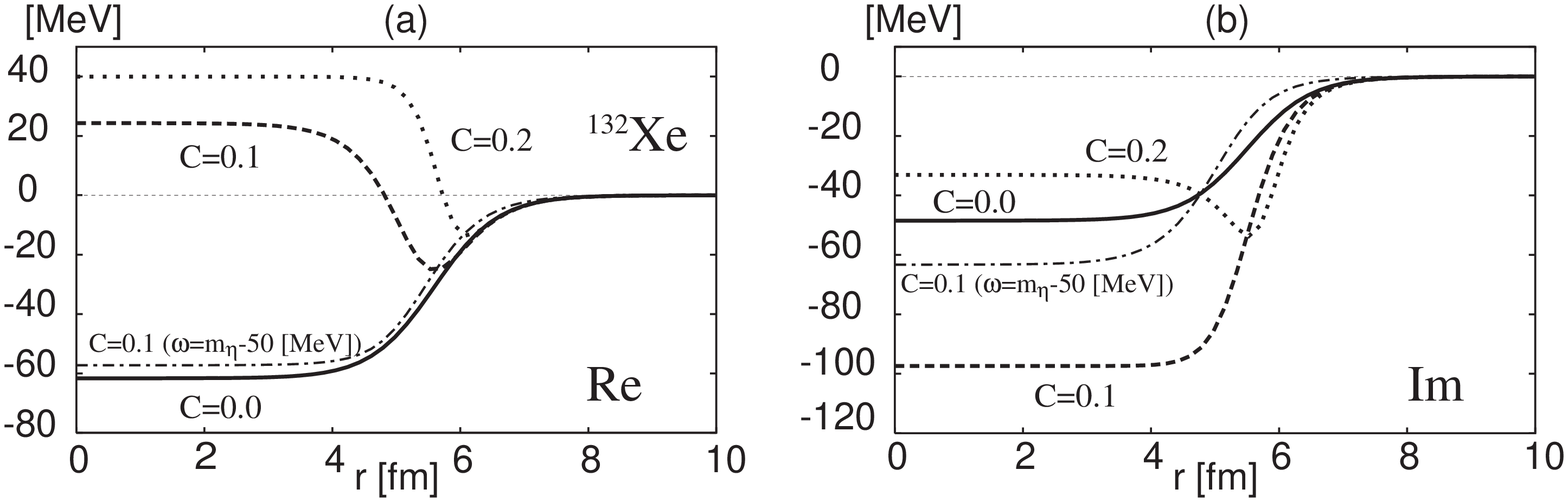}
\caption{\small The $\eta$-nucleus optical potential for $^{132}$Xe system as
a function of the radius coordinate $r$.  
The left (a) and right (b) panels show the real and
imaginary parts, respectively, for $C=0.0$ (solid line), $0.1$ (dashed
line) and $0.2$ (dotted line) with setting $\omega=m_{\eta}$.  The
dot-dashed line indicates the potential strength for $C=0.1$ with
$\omega=m_{\eta} - 50$ [MeV].
\label{potentialfig}}
\end{center}
\end{figure}

In Fig.\ref{potentialfig}, we show the $\eta$-nucleus potential for the 
$^{132}$Xe case, as an example.
In other nuclei, the potential shape is 
essentially same as the plotted one, but the radius of the repulsive core 
depends on the mass number $A$. 
As can be seen in Fig.\ref{potentialfig}, for the $C\neq 0$ cases,
the real potential turns out to be repulsive at the inner part of the
nucleus and an attractive ``pocket'' appears on the
surface.  Another interesting feature of the potential is its strong energy
dependence.  By changing the energy of the $\eta$ from $m_{\eta}$ to 
$m_{\eta}- 50$ [MeV], we find again the familiar attractive
potential even with $C=0.1$ as shown in Fig.\ref{potentialfig}.  


We investigate the level structures of  the bound states in
our optical potential.
The calculated binding energies and level  widths are summarized in Table 1
in ref.\cite{Jido:2002yb}.  
For $C \geq 0.1$ cases, the
formation of $\eta$ bound states is quite difficult 
because of both the repulsive nature of the potential inside the nucleus and 
the huge imaginary part of the potential.
For the $C=0$ case, in which no medium modifications are
included in the $N$ and $N^{*}$ properties, since the potential is
proportional to the nuclear density as we have seen in 
Fig.\ref{potentialfig}, the level structure of the bound states is similar
to that obtained in ref.\cite{hayano99}.

Although the formation of the bound states of the $\eta$ in nuclei is difficult with 
$C\sim 0.2$, which is the expected strength of the chiral restoration in the nucleus,
it would be interesting to see if the repulsive nature of the optical potential can be
observed in experiment. Here we consider the recoilless $^{12}$C(d,$^3$He) 
reaction, in which a proton is picked up from the target nucleus and 
the $\eta$ meson is left with a small momentum.

\begin{figure}[t]
\begin{center}
\epsfxsize=15.cm
\epsfbox{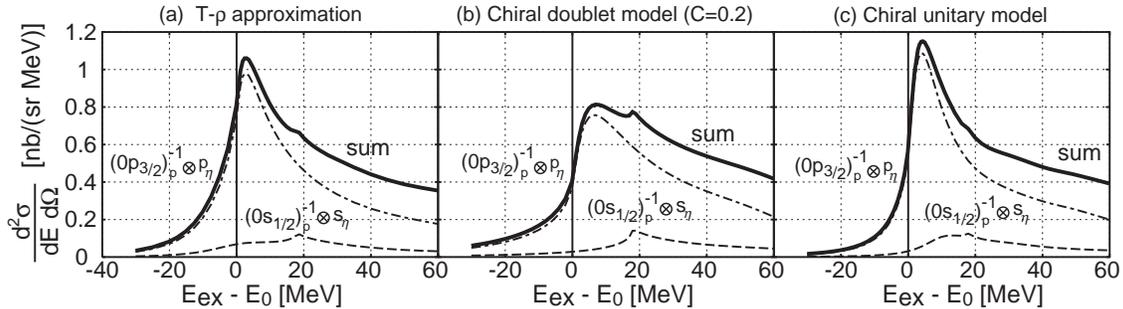}
\caption{ \small
The calculated excitation energy spectra of the $\eta$ production
in the $^{12}$C(d,$^3$He) reaction at $T_d=3.5$ [GeV] for three different
models;
(a) the $T \rho$ approximation, (b) the chiral double model with $C=0.2$, 
(c) the chiral unitary model.
The vertical line  indicates the $\eta$ production threshold energy.
\label{fig:spec}}
\end{center}
\end{figure}

The calculated spectra are shown in Fig.\ref{fig:spec} for three different models of 
the optical potential 
as functions of the excited energy which are
defined as
  $  E_{\rm ex} = m_{\eta} - B_{\eta} + (S_{p}(j_{p}) - S_{p}(p_{3/2})) $, 
where $B_{\eta}$ is the $\eta$ binding energy and $S_{p}$  the
proton separation energy.  
In Fig.\ref{fig:spec}(a), we show the result with $C=0.0$, which
corresponds to the spectrum with the $T\rho$ approximation in the optical potential.
The results with the medium corrections are shown in Fig.\ref{fig:spec}(b) 
for the $C=0.2$ case, where
the $\eta$ optical potential has the repulsive core in the center of nucleus.
In Fig.\ref{fig:spec}(c) is shown the spectrum with the optical potential obtained in the
chiral unitary model \cite{Inoue:2002xw}, where $N^*$ is dynamically generated in the 
meson-baryon scatterings.
It is seen in Fig.\ref{fig:spec}(b) that, as a result of  the repulsive nature 
of the $\eta$ potential, the whole spectrum spreads out to the 
higher energy region.
The difference of these spectra is expected to be observed in the high resolution
experiment. 

\section{Summary}

We investigate the consequences of the medium effects to $N(1535)$
$(N^*)$ through the $\eta$-mesic nuclei.
We find that sufficient reduction of the in-medium mass difference
makes the $\eta$ optical potential
repulsive at certain densities, while in the low density approximation
the optical potential is estimated to be attractive.  This leads us the
possibility of a new type of potential of the $\eta$ in nucleus
that is attractive at the surface and has a repulsive core at the center
of the nucleus.  
Unfortunately it is hard to form $\eta$ bound states in
nucleus with the expected strength of the chiral restoration in
nucleus ($C \sim 0.2$), due to the repulsive nature of the potential inside the nucleus
and its large imaginary potential.
We also evaluate the spectra of the recoilless $^{12}$C(d,$^{3}$He) reaction 
using optical potentials for different models.  The shapes of these spectra are apparently different and the 
repulsive nature in the $C=0.2$ case is seen.  
We believe that the present results are very important to investigate the
chiral nature of $N$ and $N^*$ through $\eta$ bound states.

\end{document}